\documentclass[conference]{IEEEtran}
\IEEEoverridecommandlockouts
\usepackage{cite}
\usepackage{amsmath,amssymb,amsfonts}
\usepackage{algorithmic}
\usepackage{graphicx}
\usepackage{textcomp}
\usepackage{xcolor}

\usepackage{subcaption}
\usepackage{amsmath,amsthm,amssymb}
\usepackage{floatflt}  
\usepackage{comment}

\usepackage{amsthm}
\theoremstyle{plain}

\theoremstyle{definition}

\theoremstyle{remark}

\usepackage{tikz}
\usetikzlibrary{trees}
\usepackage{caption}

\usepackage{flushend}

\begin{document}

\title{Strategic Deployment of Honeypots in Blockchain-based IoT Systems \thanks{\textcopyright 2024 IEEE. Personal use of this material is permitted. Permission from IEEE must be obtained for all other uses, in any current or future media, including reprinting/republishing this material for advertising or promotional purposes, creating new collective works, for resale or redistribution to servers or lists, or reuse of any copyrighted component of this work in other works.}}

\author{
\IEEEauthorblockN{1\textsuperscript{st} Daniel Commey}
\IEEEauthorblockA{\textit{Multidisciplinary Engineering} \\
\textit{Texas A\&M University}\\
Texas, USA\\
dcommey@tamu.edu
}
\and
\IEEEauthorblockN{2\textsuperscript{nd} Sena Hounsinou}
\IEEEauthorblockA{\textit{Computer Science \& Cybersecurity} \\
\textit{Metro State University}\\
Minnesota, USA\\
sena.houeto@metrostate.edu
}
\and
\IEEEauthorblockN{3\textsuperscript{rd} Garth V. Crosby}
\IEEEauthorblockA{\textit{Engineering Technology \& Industrial Distribution} \\
\textit{Texas A\&M University}\\
Texas, USA\\
gvcrosby@tamu.edu
}
}

\maketitle

\begin{abstract}
This paper addresses the challenge of enhancing cybersecurity in Blockchain-based Internet of Things (BIoTs) systems, which are increasingly vulnerable to sophisticated cyberattacks. It introduces an AI-powered system model for the dynamic deployment of honeypots, utilizing an Intrusion Detection System (IDS) integrated with smart contract functionalities on IoT nodes. This model enables the transformation of regular nodes into decoys in response to suspicious activities, thereby strengthening the security of BIoT networks. The paper analyses strategic interactions between potential attackers and the AI-enhanced IDS through a game-theoretic model, specifically Bayesian games. The model focuses on understanding and predicting sophisticated attacks that may initially appear normal, emphasizing strategic decision-making, optimized honeypot deployment, and adaptive strategies in response to evolving attack patterns.

\end{abstract}


\section{Introduction}

The Internet of Things (IoT) has revolutionized numerous sectors through its vast network of interconnected devices. However, the rapid expansion of the IoT introduces significant security and privacy concerns, as these devices often have limited computational resources, making them vulnerable to attacks \cite{alajlan_cybersecurity_2023}. Traditional security solutions, which sometimes rely on centralized systems, encounter scalability issues and introduce further vulnerabilities \cite{alajlan_cybersecurity_2023, erfan_game-theoretic_2022}.

Blockchain technology has emerged as a promising solution to enhance the security and privacy of IoT systems. Its decentralized nature facilitates secure, transparent, and tamper-resistant transactions, effectively mitigating the challenges associated with the IoT's growth \cite{alajlan_cybersecurity_2023, erfan_game-theoretic_2022, sandner_convergence_2020}. Integrating blockchain and IoT, offers a more secure and resilient framework for IoT systems. However, BIoT systems still face unique cybersecurity challenges, necessitating the development of innovative defense mechanisms \cite{alajlan_cybersecurity_2023}.

Honeypots have proven to be an effective tool in the fight against cyber threats. By acting as decoys, honeypots attract and analyze cyberattacks, providing valuable insights into the methods and strategies employed by attackers. The strategic deployment of honeypots in BIoT systems strengthens the overall security posture and addresses pressing privacy concerns \cite{ul_hassan_privacy_2019}. 

Numerous studies have explored using honeypots in various settings, leveraging game theory to optimize their effectiveness. La et al. \cite{la_deceptive_2016} and Diamantoulakis et al. \cite{diamantoulakis_game_2020} investigated the deployment of honeypots in IoT networks and smart grid systems to counteract DDoS attacks using Bayesian games of incomplete information. Li et al. \cite{li_honeypot-enabled_2021} proposed using honeypots in smart grids to monitor cyber penetration, applying stochastic games to understand the complex dynamics of these environments. 

In the context of Advanced Persistent Threats (APTs), Tian et al. \cite{tian_honeypot_2021} introduced prospect theoretic games to assess bounded rational behavior when deploying honeypots in Software-Defined Networks (SDNs) within the Industrial Internet of Things (IIoT). Wang et al. \cite{wang_strategic_2017} focused on honeypot deployment in Advanced Metering Infrastructure (AMI) networks to strategize against DDoS attacks using Bayesian games. 

Shi et al. \cite{shi_research_2021} contributed a three-party evolutionary game model for array honeypot systems, introducing a novel aspect to honeypot strategy. Carroll \cite{carroll_game_2011} reviewed game-theoretic approaches in cybersecurity, highlighting the role of deception in securing networks through signaling and Bayesian repeated games. Li et al. \cite{li_game-theoretic_2019} explored distributed honeypot schemes and game theory models like Bayesian and evolutionary games to address network security challenges in IoT, Unmanned Aerial Vehicles (UAVs), and cloud computing. 

Boumkheld et al. \cite{boumkheld_honeypot_2019} focused on honeypots in smart grids for Security Information and Event Management (SIEM) training and post-incident analysis, applying a Bayesian game with complete but imperfect information. Tian et al. \cite{tian_honeypot_2019} and Lee et al. \cite{lee_you_2022} explored honeypot applications in Cyber-Physical Systems (CPS) and industrial control systems, proposing game-theoretical models to optimize defense against APTs, considering honeypot allocation costs and human analysis. Florea and Craus \cite{florea_game-theoretic_2022} reviewed network security tactics, including honeypots in software-defined networks, discussing cyber deception games and honeypots as cyber camouflage methods.

Despite the extensive research on honeypot deployment in various domains, there remains a gap in the literature regarding the strategic deployment of honeypots in BIoT systems. The unique characteristics of BIoT environments, such as the decentralized architecture and smart contracts, necessitate the development of tailored defense mechanisms that can adapt to the evolving threat landscape.

We propose an AI-powered model for strategically deploying honeypots in BIoT systems to address this gap. Our model leverages game-theoretic principles to optimize defense mechanisms against sophisticated cyber threats, contributing to the field with a deceptive defense solution designed to secure BIoT environments and enhance their resilience.

The main contributions of this paper are as follows:

\begin{enumerate}
    \item We introduce an AI-powered system model for the dynamic deployment of honeypots in BIoT systems, utilizing an Intrusion Detection System (IDS) integrated with smart contract functionalities on IoT nodes.
    
    \item We develop a game-theoretic model, specifically Bayesian games, to analyze the strategic interactions between potential attackers and the AI-enhanced IDS, focusing on understanding and predicting sophisticated attacks that may initially appear normal.
    
    \item We conduct simulations to evaluate the effectiveness of various honeypot deployment strategies in a BIoT security context, considering the trade-off between detecting attacks and minimizing false positives.
    
    \item We provide insights and recommendations for developing more sophisticated intrusion detection systems that optimize the balance between security and operational costs in BIoT environments.
\end{enumerate}

The remainder of this paper is organized as follows: Section \ref{sec:proposed_model} outlines our proposed AI-powered system model for the dynamic deployment of honeypots in BIoT systems; Section \ref{sec:game_model} discusses the game-theoretic model and its analysis; Section \ref{sec:simulation} presents the simulation setup and results; Section \ref{sec:conclusion} concludes the paper and outlines future research directions.

\section{Proposed Model}
\label{sec:proposed_model}

Blockchain-based Internet of Things (BIoT) networks comprise interconnected smart devices crucial for service delivery. These devices, including smart home appliances, health monitors, and environmental sensors, perform data collection and real-time monitoring autonomously through smart contracts. Their critical role and blockchain connectivity make them attractive targets for cyber attackers.

We propose a multi-layered security strategy incorporating honeypots as a proactive defense mechanism to address these security challenges. Inspired by the work of La et al. \cite{la_deceptive_2016}, our model integrates an AI-powered Intrusion Detection System (IDS) with smart contract capabilities to scrutinize incoming network traffic. Suspicious traffic is rerouted to honeypots for analysis, while legitimate traffic continues uninterrupted, ensuring continuous service.

\subsection{System Architecture}

\begin{figure}[htbp]
\centering
\includegraphics[width=0.5\textwidth]{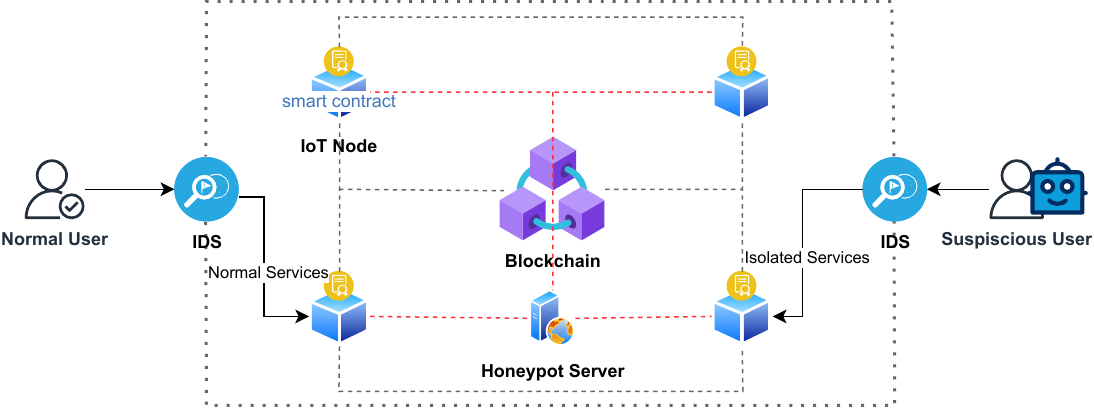}
\caption{System Model of BIoTs with IDS}
\label{fig:biots_ids_model}
\end{figure}

As depicted in Figure \ref{fig:biots_ids_model}, the IDS dynamically changes the status of IoT nodes within the system, isolating them from the main network when suspicious activity is detected. This isolation allows suspicious traffic to be redirected to a honeypot server, which can be safely analyzed without compromising the main network's integrity. Meanwhile, normal traffic continues to flow seamlessly, ensuring uninterrupted service.

The proposed system architecture consists of the following key components:

\begin{enumerate}
    \item \textbf{IoT Nodes:} These are the smart devices connected to the BIoT network, performing various tasks such as data collection, monitoring, and actuation.
    
    \item \textbf{Blockchain Network:} The BIoT system is built on a blockchain network that ensures secure, transparent, and tamper-proof transactions among the IoT nodes. Smart contracts are deployed on the blockchain to facilitate the autonomous execution of predefined rules and actions.
    
    \item \textbf{AI-powered IDS:} The Intrusion Detection System is a critical component of the proposed model. It leverages machine learning algorithms to analyze network traffic patterns and detect suspicious activities. The IDS is trained on a diverse dataset of normal and malicious traffic patterns to enhance its accuracy and effectiveness.
    
    \item \textbf{Honeypot Server:} The honeypot server is a dedicated system that hosts various honeypots designed to attract and trap attackers. When the IDS detects suspicious traffic, it redirects it to the honeypot server for further analysis. The honeypots mimic vulnerabilities and entice attackers to reveal their tactics and intentions.
    
    \item \textbf{Smart Contracts:} Smart contracts play a vital role in automating the deployment of honeypots and the isolation of compromised nodes. When the IDS detects a potential threat, it triggers a smart contract that automatically reconfigures the network topology, isolating the suspicious node and redirecting its traffic to the honeypot server.
\end{enumerate}

\subsection{Workflow}

The workflow of the proposed model can be summarized as follows:

\begin{enumerate}
    \item The AI-powered IDS continuously monitors the network traffic of each IoT node in the BIoT system.
    
    \item When the IDS detects suspicious traffic patterns, it triggers a smart contract on the blockchain network.
    
    \item The smart contract automatically isolates the suspicious node from the main network and redirects its traffic to the honeypot server.
    
    \item The honeypot server analyzes the redirected traffic to identify the attacker's tactics, techniques, and procedures (TTPs).
    
    \item Based on the analysis, the system updates its threat intelligence and refines the IDS's detection models to improve its accuracy and effectiveness.
    
    \item The isolated node remains in the honeypot environment until it is deemed safe to reintegrate into the main network.
    
    \item Normal traffic continues to flow uninterrupted, ensuring the availability and functionality of the BIoT system.
\end{enumerate}

Integrating smart contracts with an AI-powered IDS enables our BIoT security model to leverage smart contracts' self-executing capabilities for enforcing and automating security protocols. Smart contracts enable decentralized and autonomous actions, such as real-time rerouting of suspicious network traffic to honeypots, based on the AI's analysis of data patterns and anomalies. The AI component employs machine learning algorithms to continuously analyze network traffic patterns and compare them against known threat signatures and anomalies, while smart contracts ensure the automatic deployment of honeypots upon anomaly detection.

The proposed model offers several advantages over traditional security approaches. By dynamically isolating suspicious nodes and redirecting their traffic to honeypots, the system can effectively contain potential threats and prevent them from spreading across the network. Smart contracts enable automated and rapid response to detected anomalies, reducing the response time and minimizing the potential impact of an attack. Furthermore, the AI-powered IDS continuously learns and adapts to new threat patterns, enhancing the system's resilience against evolving attack vectors.

One of the key advantages of our proposed model is its dynamic honeypot deployment strategy. Traditional honeypot deployment approaches often introduce additional computational overhead and resource consumption by dedicating specific nodes as permanent honeypots, which may impact the overall performance of the BIoT system and increase operational costs. In contrast, our approach dynamically changes the state of normal nodes to honeypots when suspicious activity is detected. This dynamic adaptation allows for more efficient resource utilization, as the nodes can continue to perform their regular tasks when not acting as honeypots. By temporarily isolating suspicious nodes and redirecting their traffic to the honeypot server, our model minimizes the performance impact on the overall BIoT system while still providing effective threat detection and analysis capabilities. This dynamic approach also leads to cost savings, as it eliminates the need for dedicated honeypot hardware and reduces the overall resource consumption of the security infrastructure.

However, the proposed model also presents some challenges. The accuracy and effectiveness of the IDS heavily depend on the quality and diversity of the training dataset. Obtaining a comprehensive dataset that covers a wide range of attack scenarios can be difficult and time-consuming. Additionally, the dynamic nature of the honeypot deployment raises questions about the optimal strategies for both defenders and attackers. Defenders must decide when and where to deploy honeypots based on the perceived threat level, while attackers may adapt their strategies to avoid detection and maximize their chances of success.

To address these challenges, we propose a game-theoretic approach to model the strategic interactions between the defenders (i.e., the BIoT system equipped with the AI-powered IDS) and the attackers. Game theory provides a mathematical framework for analyzing the decision-making processes of rational agents in strategic situations. By modeling the incentives, costs, and benefits of different strategies, game theory can help us derive optimal defense mechanisms that consider the attackers' potential actions.

In our context, the game-theoretic analysis will focus on the following challenges:

\begin{enumerate}
    \item Optimal honeypot deployment: Determining the best strategies for deploying honeypots based on the perceived threat level and the cost-benefit trade-offs.
    \item Adaptive defense mechanisms: Developing defense strategies that can adapt to the evolving strategies of the attackers, taking into account the potential for sophisticated attacks that may initially appear normal.
    \item Balancing security and performance: Identifying the optimal balance between maximizing the security of the BIoT system and minimizing the performance overhead introduced by the honeypot deployment and IDS operations.
\end{enumerate}

\section{Game Model}
\label{sec:game_model}

\subsection{Scenario Description}
We analyze the dynamics between defenders and potential attackers in a BIoT system. Attackers aim to exploit these networks, employing tactics from naive, basic methods to sophisticated strategies designed to mimic legitimate operations. Sophisticated attackers, in particular, present a significant challenge; their deep understanding of the BIoT environment allows them to carefully plan their activities to avoid detection, often behaving indistinguishably from legitimate users to infiltrate the system. Defenders, equipped with AI-powered IDS and smart contracts, monitor network activities to distinguish between legitimate operations and potential threats, deploying honeypots as a strategic countermeasure to entrap attackers. The IDS's activity assessment influences the deployment decision, with a higher probability of honeypot use as the risk of sophisticated attack increases.

In the BIoT security game, defenders and attackers have distinct actions. Defenders can either deploy a honeypot \((D_h)\), aiming to trap attackers by simulating vulnerabilities or choose not to \((D_n)\). Attackers, on the other hand, decide to either launch an attack \((A_t)\) or abstain \((A_b)\). Sophisticated attackers might attack discreetly, while naive attackers are more likely to be readily detected. The decision to abstain reflects a strategic choice to avoid engagement under potentially risky conditions. For instance, sophisticated attackers, appearing as normal users, might attack with only a 50\% probability to minimize suspicion, whereas naive attackers, easily identified as suspicious, may attack with a 100\% probability due to their less cautious approach.

\subsection{One-shot Game}

We model the attack and defense scenario in the BIoT system as a one-shot, sequential game between an attacker (A) and a defender (D), concluding after a single round of actions \cite{la_deceptive_2016}.

Table \ref{tab:symbols} presents the notation used in the game model to clarify the symbols and their meanings for the reader.

\begin{table}[htbp]
\centering
\footnotesize
\caption{Notation for the game model}
\label{tab:symbols}
\begin{tabular}{p{0.1\linewidth}p{0.82\linewidth}}
\hline
Symbol & Description \\ \hline
$D_a$ & Defender actions: $\{D_h, D_n\}$, deploy or not deploy honeypot \\
$A_a$ & Attacker actions: $\{A_t, A_b\}$, attack or abstain \\
$\theta_A$ & Attacker types: $\{Sophisticated, Naive\}$ \\
$C_h$ & Cost of deploying a honeypot \\
$C_{a,s}$ & Cost to defender for a successful attack by sophisticated attacker (no honeypot) \\
$C_{a,n}$ & Cost to defender for a successful attack by naive attacker (no honeypot) \\
$B_{a,s}$ & Benefit to sophisticated attacker from a successful attack \\
$B_{a,n}$ & Benefit to naive attacker from a successful attack \\
$C_{d,s}$ & Cost to sophisticated attacker for being detected (attacking honeypot) \\
$C_{d,n}$ & Cost to naive attacker for being detected (attacking honeypot) \\
$B_{d,s}$ & Benefit to defender for detecting a sophisticated attacker \\
$B_{d,n}$ & Benefit to defender for detecting a naive attacker \\
$P_n$ & Penalty for misidentifying a normal user \\
$p(S)$ & Probability that the attacker is sophisticated \\ \hline
\end{tabular}
\end{table}

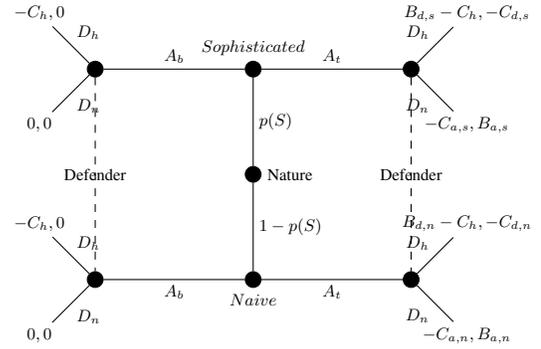
\begin{figure}[htbp]
\centering
  \small
  \begin{tikzpicture}[scale=0.70, every node/.style={transform shape}, node distance=1.5cm]
    \tikzstyle{action}=[draw,circle,fill]; \tikzstyle{end}=[]; \node
    (0) [action,label=right:Nature] at (0,0) {};

    \node (1) [action,label=above:$Sophisticated$] at (0,2) {}; \node (2)
    [action,label=below:$Naive$] at (0,-2) {};

    \node (3) [action] at (3,2) {};
     \node (3a) [end,above right of=3] {$B_{d,s} - C_h, -C_{d,s}$} ;
     \node (3b) [end,below right of=3] {$-C_{a,s}, B_{a,s}$} ;

    \node (4) [action] at (-3,2) {};
     \node (4a) [end,above left of=4] {$-C_h, 0$} ;
     \node (4b) [end,below left of=4] {$0, 0$} ;

    \node (5) [action] at (3,-2) {};
     \node (5a) [end,above right of=5] {$B_{d,n} - C_h, -C_{d,n}$} ;
     \node (5b) [end,below right of=5] {$-C_{a,n}, B_{a,n}$} ;

    \node (6) [action] at (-3,-2) {};
     \node (6a) [end,above left of=6] {$-C_h, 0$} ;
     \node (6b) [end,below left of=6] {$0, 0$} ;

    \path (0) edge node[right] {$p(S)$} (1) edge node[right] {$1-p(S)$} (2)
    (4) edge [dashed] node[] {Defender} (6) (3) edge [dashed] node[]
    {Defender} (5) (1) edge node[above] {$A_t$} (3) edge node[above]
    {$A_b$} (4) (2) edge node[below] {$A_t$} (5) edge node[below] {$A_b$}
    (6) (3) edge node[above left] {$D_h$} (3a) edge node[below left]
    {$D_n$} (3b) (4) edge node[above right] {$D_h$} (4a) edge node[below
    right] {$D_n$} (4b) (5) edge node[above left] {$D_h$} (5a) edge
    node[below left] {$D_n$} (5b) (6) edge node[above right] {$D_h$} (6a)
    edge node[below right] {$D_n$} (6b);
  \end{tikzpicture}
  
  \caption{Game tree for the BIoTs Game Model}
\label{fig:biots_game_tree}
\end{figure}

Figure \ref{fig:biots_game_tree} depicts the game tree for the BIoT security game. The game begins with Nature deciding the attacker's type, introducing uncertainty. The attacker's behavior is categorized as either Sophisticated (\(S\)) or Naive (\(N\)). To capture this uncertainty, the game defines two primary information sets: \(N = \{(N | S), (N | N)\}\) and \(S = \{(S | S), (S | N)\}\). In these sets, the first element represents the defender's perception based on the observed signal, while the second element denotes the actual type of the attacker.

\subsection{Game Formulation}
We formulate the game as a Bayesian game to capture the uncertainty regarding the attacker's type. The defender must decide whether to deploy a honeypot based on the perceived threat level, influenced by the IDS's accuracy. Attackers choose their actions based on their type and perceived defensive measures.

Table \ref{tab:payoffs} presents the game's payoff matrix, detailing each player's utilities under different action profiles and attacker types.

\begin{itemize}
    \item \textbf{Players:} The Defender ($D$) and the Attacker ($A$). The attacker can be either sophisticated ($S$) or naive ($N$), formally captured as a type $\theta \in \{S, N\}$.
    
    \item \textbf{Actions:} 
    \begin{itemize}
        \item Defender: Deploy honeypot ($D_h$) or not ($D_n$). Formally, the action set is $D_a = \{D_h, D_n\}$.
        \item Attacker: Attack ($A_t$) or abstain ($A_b$). The action set is $A_a = \{A_t, A_b\}$.
    \end{itemize}
    
    \item \textbf{Payoffs:} $U_i(\theta, a_D, a_A)$, where $i$ stands for player type (defender or attacker).
\end{itemize}

\begin{table*}[htbp]
\centering
\footnotesize
\caption{Payoff Structure of the Game}
\label{tab:payoffs}
\begin{tabular}{ccccc}
\hline
 & \multicolumn{2}{c}{Sophisticated Attacker} & \multicolumn{2}{c}{Naive Attacker} \\
Defender / Attacker & Attack ($A_t$) & Abstain ($A_b$) & Attack ($A_t$) & Abstain ($A_b$) \\ \hline
Deploy Honeypot ($D_h$) & $(B_{d,s} - C_h, -C_{d,s})$ & $(-C_h, 0)$ & $(B_{d,n} - C_h, -C_{d,n})$ & $(-C_h, 0)$ \\
Not Deploy ($D_n$) & $(-C_{a,s}, B_{a,s})$ & $(0, 0)$ & $(-C_{a,n}, B_{a,n})$ & $(0, 0)$ \\ \hline
\end{tabular}
\end{table*}

\subsection{Equilibrium Analysis}
We define the Bayesian Nash Equilibrium (BNE) concept to analyze the game's equilibria.

\textbf{Definition (BNE):}
A strategy profile $(\sigma^{*}, \tau^{*})$, accompanied by a belief system $\mu^{*}$, forms a BNE if, under $\mu^{*}$, $\sigma^{*}$ and $\tau^{*}$ represent the best response strategies for the attacker and defender, respectively.

\subsubsection{Pure Strategy BNE}
In the BIoT game, a pure strategy BNE arises when the defender opts to deploy a honeypot ($\tau^{*} = D_h$) with probability 1 if the expected utility of such deployment exceeds that of non-deployment, factoring in both the cost of honeypot deployment and the benefits of attacker detection.

\textit{Proof:}
The defender's deployment decision hinges on whether the expected utility of deploying a honeypot surpasses abstaining based on beliefs $\mu$ about the attacker's type. The defender prefers deployment ($D_h$) over non-deployment ($D_n$) if:
\begin{equation}
C_h < p_S p_{A_t|S} (B_{d,s} + C_{a,s}) + (1 - p_S) p_{A_t|N} (B_{d,n} + C_{a,n}),
\end{equation}
indicating that honeypot deployment is the defender's best strategy under these conditions.

\subsubsection{Mixed Strategy BNE}
A mixed-strategy BNE, where both parties randomize their strategies, emerges when neither can justify a pure strategy based on expected costs and benefits alone.

\textit{Proof:}
For a mixed-strategy BNE, we consider the defender deploying a honeypot with probability $\beta$ and the attacker choosing to attack based on their type with probabilities $\alpha_S$ and $\alpha_N$. Equilibrium requires indifference in strategy choice for both parties, leading to equilibrium probabilities $\beta$, $\alpha_S$, and $\alpha_N$, contingent on the game's parameters.
To achieve a mixed-strategy BNE, the following indifference conditions must hold:

Defender's Indifference:
The defender must be indifferent between deploying $(D_h)$ and not deploying $(D_n)$:
\begin{equation}
EU_{D,h}(\beta) = EU_{D,n}(\beta). 
\end{equation}

Attackers' Indifference:
Each attacker type must be indifferent between attacking $(A_t)$ and abstaining $(A_b)$:
\begin{align}
EU_{A,S|A_t}(\alpha_S) &= EU_{A,S|A_b}(\alpha_S) \quad \text{(sophisticated)},\\
EU_{A,N|A_t}(\alpha_N) &= EU_{A,N|A_b}(\alpha_N) \quad \text{(naive)}. 
\end{align}

Expanding these expected utilities in terms of the game's parameters and solving the resulting system of equations for  $\beta$, $\alpha_S$, and $\alpha_N$ will yield the mixed-strategy BNE. For a detailed mathematical framework for analyzing Bayesian Nash Equilibria, refer to \cite{fudenberg_game_1991}.

\section{Simulation}
\label{sec:simulation}

We conducted a series of simulation experiments to evaluate the effectiveness of various honeypot deployment strategies in a BIoT security context. The objective was to optimize the balance between detecting attacks (both naive and sophisticated) and minimizing the misidentification of legitimate traffic as malicious (false positives).

The simulation parameters, as shown in Table \ref{tab:simulation_parameters}, were carefully chosen to reflect the cost-benefit analysis of honeypot deployment, the likelihood of facing sophisticated or naive attackers, and the performance of the IDS system in detecting these attacks. The rationale behind each parameter value is provided to ensure the simulation aligns with realistic scenarios.

\begin{table}[htbp]
\centering
\footnotesize
\caption{Simulation Parameters}
\label{tab:simulation_parameters}
\begin{tabular}{p{0.2\linewidth}p{0.09\linewidth}p{0.58\linewidth}}
\hline
Parameter & Value & Rationale \\
\hline
\(C_{a,s}\), \(C_{a,n}\) & 50, 30 & Attacks by sophisticated actors likely have a higher impact, thus the higher cost. Naive attackers are presumed to cause less damage. \\
\(B_{d,s}\), \(B_{d,n}\) & 10, 8 & Detecting a sophisticated attack may be more valuable due to their potentially higher impact and subtlety. \\
\(C_h\) & 2 & Honeypot deployment should be relatively inexpensive compared to attack costs. \\
\(P_n\) & 1 & This specific value allows for exploring the impact of different penalty levels for false positives. \\
\(p_{sophisticated}\) & 0.4 & Indicates the likelihood that a given 'attack' is from a sophisticated actor. \\
\(p_{attack\_soph}\), \(p_{attack\_naive}\) & 0.5, 0.9 & Naive attackers are modeled as more likely to attack than sophisticated ones, reflecting less cautious behavior. \\
\(TPR_{soph}\), \(TPR_{naive}\) & 0.6, 0.9 & Reflects the notion that the IDS is better at detecting unsophisticated attacks, which are less complex. \\
\(FPR\) & 0.05 & A fairly low false positive rate is used for this simulation to represent a well-tuned IDS. \\
\hline
\end{tabular}
\end{table}

The utility calculation for each event in the simulation was formulated as follows:

Let \( U_D(e) \) be the defender's utility for an event \( e \), which can be a legitimate event, a sophisticated attack, or a naive attack.

For a legitimate event:
\begin{equation}
U_D(e) = 
\begin{cases}
-P_n, & \text{if honeypot is deployed} \\
     & \text{(false positive)}, \\
0,   & \text{otherwise}.
\end{cases}
\end{equation}

For a sophisticated attack:
\begin{equation}
U_D(e) = 
\begin{cases}
B_{d,s} - C_h, & \text{if honeypot is deployed} \\
              & \text{and attack detected}, \\
-C_h,         & \text{if honeypot is deployed} \\
              & \text{but attack not detected}, \\
-C_{a,s},     & \text{if honeypot is not deployed} \\
              & \text{(attack successful)}, \\
0,            & \text{if honeypot is not deployed} \\
              & \text{and no attack occurs}.
\end{cases}
\end{equation}

For a naive attack:
\begin{equation}
U_D(e) = 
\begin{cases}
B_{d,n} - C_h, & \text{if honeypot is deployed} \\
              & \text{and attack detected}, \\
-C_h,         & \text{if honeypot is deployed} \\
              & \text{but attack not detected}, \\
-C_{a,n},     & \text{if honeypot is not deployed} \\
              & \text{(attack successful)}, \\
0,            & \text{if honeypot is not deployed} \\
              & \text{and no attack occurs}.
\end{cases}
\end{equation}

The average utility for a strategy over a trial was calculated as:

\begin{equation}
\overline{U_D} = \frac{1}{N} \sum_{i=1}^{N} U_D(e_i),
\end{equation}

where \( N \) is the total number of events in the trial.

\subsection{Simulation Setup and Implementation}
The simulation began by initializing a set of honeypot deployment strategies, including fixed threshold strategies (FS50, FS60, FS70, FS80, FS90) and a variable strategy (VS) that adapted the deployment probability based on the IDS's F1-score. Each simulated trial represented a day of network operation with a mix of legitimate activity and a smaller, defined percentage of attacks from both sophisticated and naive types.

The simulation iterated through the events, categorizing them and determining the deployment strategy for each event based on the IDS's F1-score. Fixed strategies deployed a honeypot if the F1-score exceeded a set threshold, while the variable strategy adjusted the deployment probability dynamically.

The utility was calculated for each event to assess whether an attack was detected or missed, and the average utility achieved by each strategy was determined at the end of each trial.

\subsection{Simulation Experiments and Results}

We conducted three main simulation experiments to evaluate the effectiveness of the proposed honeypot deployment strategies:

\begin{enumerate}
    \item Comparison of deployment strategy utilities with and without false positive penalties (Figure \ref{fig:strategy_utilities_comparison}): This experiment assessed the performance of each strategy in scenarios where false positives were penalized and not penalized. The variable strategy (VS) consistently performed better, suggesting its effectiveness in minimizing false positives and associated costs.

    \begin{figure}[htbp]
    \centering
    \includegraphics[width=\linewidth]{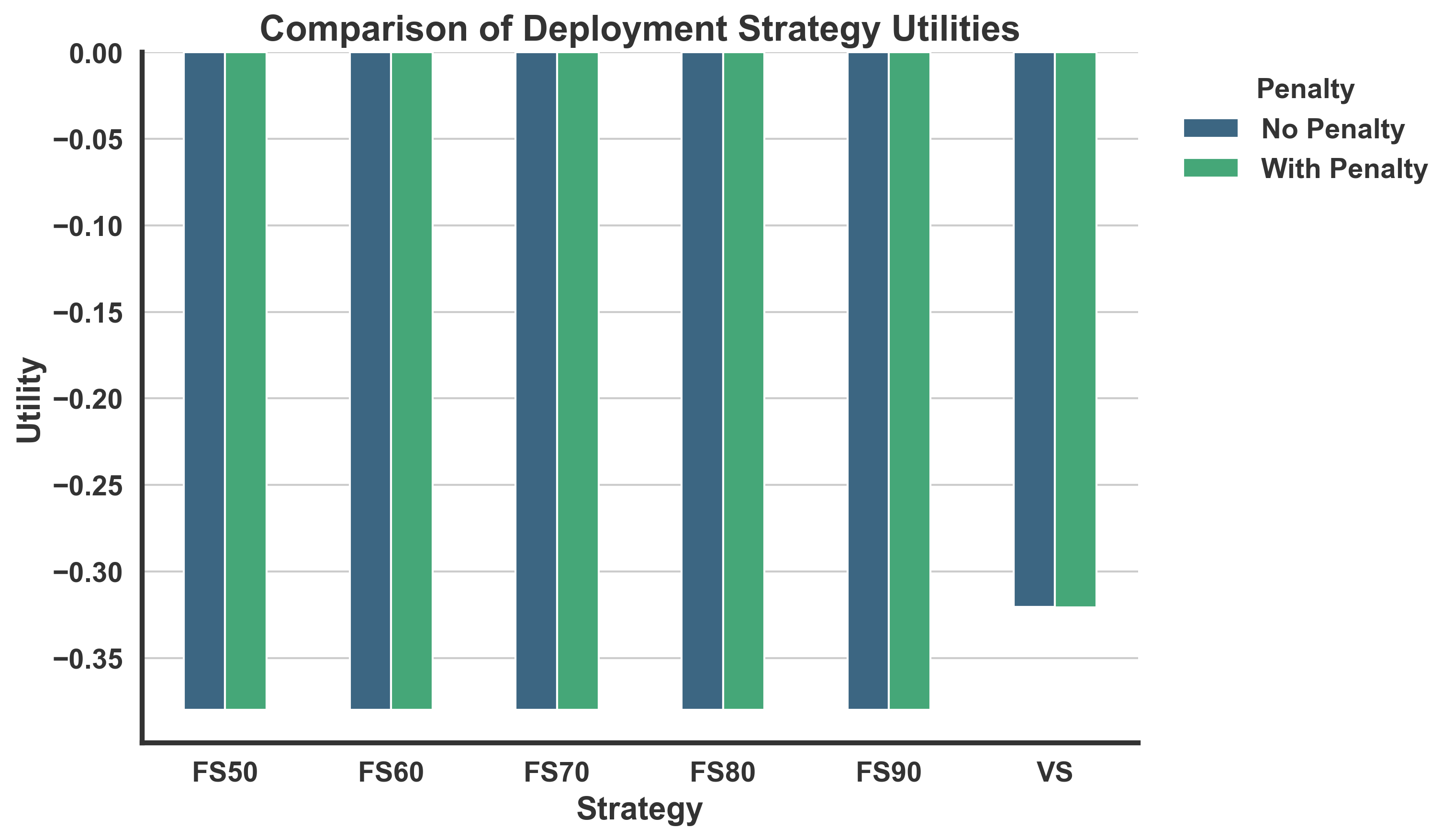}
    \caption{Comparison of Deployment Strategy Utilities without and with False Positive Penalties}
    \label{fig:strategy_utilities_comparison}
    \end{figure}

    \item Impact of varying cost of deployment and penalties on strategy utility (Figure \ref{fig:strategy_utilities_varying}): This experiment explored the effects of varying the cost of honeypot deployment and the penalties for false positives on the utility of each strategy. The results showed that increasing deployment costs reduced utility, but the strategies demonstrated resilience to cost changes.

    \begin{figure}[htbp]
    \centering
    \includegraphics[width=\linewidth]{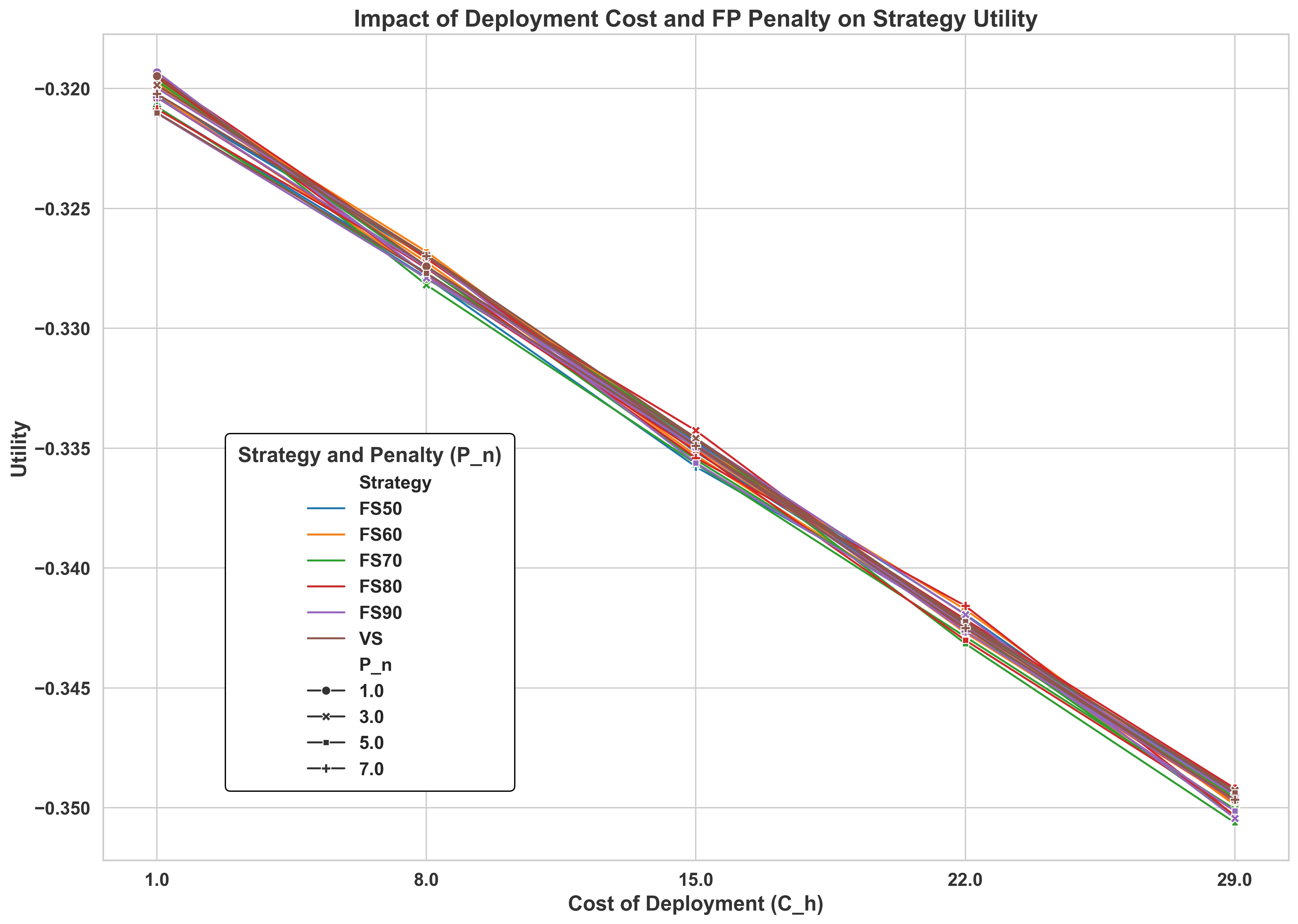}
    \caption{Impact of Varying Cost of Deployment and Penalties on Strategy Utility}
    \label{fig:strategy_utilities_varying}
    \end{figure}

    \item Comparison of deployment strategy utilities with varying attack rates (Figure \ref{fig:strategy_utilities_attackrate}): This experiment evaluated the performance of the strategies under different attack rates. The variable strategy exhibited a stable, gradual decrease in utility as the attack rate increased, suggesting its robustness to variations in attack rates, while the fixed strategies showed inconsistent utility fluctuations.

    \begin{figure}[htbp]
    \centering
    \includegraphics[width=\linewidth]{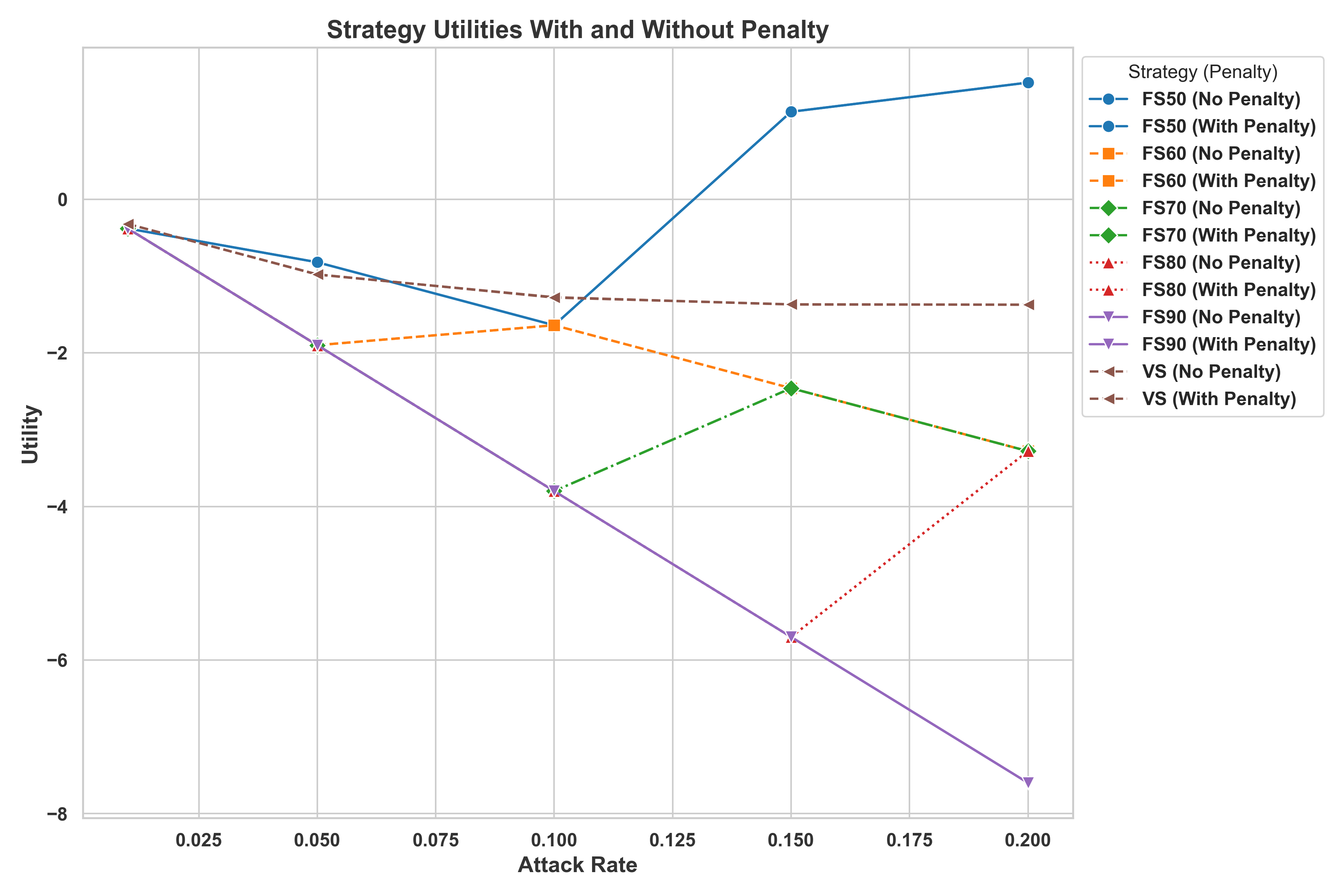}
    \caption{Comparison of Deployment Strategy Utilities with Attack Rate}
    \label{fig:strategy_utilities_attackrate}
    \end{figure}
    
\end{enumerate}

These simulation experiments provide insights into the effectiveness of different honeypot deployment strategies in various BIoT security scenarios. They highlight the importance of selecting the appropriate defense strategy based on the specific operational environment and the potential costs associated with security breaches and false alarms. The results demonstrate the effectiveness of our proposed AI-powered dynamic honeypot deployment approach in optimizing the balance between security and operational efficiency in BIoT systems.

\section{Discussion and Conclusion}
\label{sec:conclusion}

This paper introduced an AI-powered model for strategically deploying honeypots in Blockchain-based Internet of Things (BIoT) systems. By leveraging game-theoretic principles and dynamic adaptation based on the perceived threat level, our approach enhances the security and resilience of BIoT systems while minimizing the performance overhead associated with traditional honeypot deployment strategies.

The game-theoretic analysis provided valuable insights into the strategic interactions between defenders and attackers, revealing the conditions under which pure and mixed strategy equilibria emerge. The simulation results demonstrated the effectiveness of our proposed AI-powered dynamic honeypot deployment approach in optimizing the balance between security and operational efficiency, particularly when compared to fixed deployment strategies.

However, our study has some limitations that should be acknowledged. First, while based on carefully chosen parameters and rationale, the simulation experiments may not fully capture the complexity and diversity of real-world BIoT systems. Future research could involve more comprehensive simulations or real-world experiments to validate the performance of our proposed approach in more complex and realistic settings.

Second, the game-theoretic model assumes that attackers and defenders are rational agents with complete information about the game's parameters. Attackers and defenders may have bounded rationality and incomplete information, which could impact their decision-making processes. Future work could explore more complex game-theoretic models, such as Bayesian games with incomplete information or evolutionary games, to capture these real-world considerations.

Third, while the simulation results provide valuable insights into the effectiveness of different honeypot deployment strategies, they are based on specific parameter settings and assumptions. Further sensitivity analysis and robustness tests could be conducted to assess the proposed approach's performance under various parameter configurations and assumptions.

Despite these limitations, our study makes significant contributions to BIoT security. By introducing an AI-powered model for the strategic deployment of honeypots and analyzing the problem through a game-theoretic lens, we provide a foundation for developing more sophisticated and adaptive defense mechanisms in BIoT systems.

Future research directions could include incorporating more advanced machine learning techniques, such as deep reinforcement learning, to enable the AI-powered IDS to continuously adapt its detection models and honeypot deployment strategies based on real-time feedback and evolving threat patterns; investigating the integration of our proposed approach with other security mechanisms, such as blockchain-based access control or secure multi-party computation, to create a more comprehensive defense framework for BIoT systems; exploring the scalability and performance of the proposed approach in large-scale BIoT networks with a high volume of devices and traffic; and conducting user studies to assess the usability and acceptability of our proposed approach from the perspective of BIoT system administrators and end-users.

\section*{Acknowledgment}
This paper was presented at the 6th IEEE International Conference on Artificial Intelligence Circuits and Systems (IEEE AICAS 2024).

\bibliographystyle{IEEEtran} 
\bibliography{references} 

\end{document}